\documentclass[prb,twocolumn,showpacs,preprintnumbers]{revtex4}
\usepackage{graphicx}
\usepackage{dcolumn}

\begin{document}

\title{On themomagnetic effects due to the supeconducting
fluctuations:\\ Reply to arXiv:1012.4361 by  Serbyn, Skvortsov,
and Varlamov }

\author{ A. Sergeev}
\email{asergeev@eng.buffalo.edu} \affiliation{ Research
Foundation, University at Buffalo, Buffalo, New York 14260}
\author{M. Yu. Reizer} \affiliation{5614 Naiche Rd.
Columbus, Ohio 43213}
\author{V. Mitin} \affiliation{Electrical Engineering
Department, University at Buffalo, Buffalo, New York 14260}
%\date{\today}

\begin{abstract}
In a recent Reply [1] to our Comment [2], Serbyn, Skvortsov, and
Varlamov raised a question of microscopic description, which we
did not touch in [2], and criticized our work [3]. They hopefully
agreed with one of our key result [3] that the effective heat
current vertex for fluctuating Cooper pairs (the Aslamzov-Larkin
block in the diagram technique) is modified in the magnetic field,
so "the heat current is proportional to the gauge-invariant
momentum" [1] . However, they stated that in [3] we have
overlooked the same correction to the electric current vertex and
in this way we lost their huge thermomagnetic effect that does not
require any particle-hole asymmetry and, therefore, prevails over
the ordinary thermomagnetic effects by at least {\it five orders
in magnitude} in ordinary superconductors near $T_c$ and strongly
dominates in the temperature range up to $\sim 100 T_c$. Here we
address their criticism  with all details and show that our
calculations in [3] are correct.

\end{abstract} \maketitle

In our Comment [2] on "Giant Nernst Effect due to Fluctuating
Cooper Pairs in Superconductors" [4], we highlighted that the
magnetization currents do not transfer the heat. Therefore, the
amendment of the heat current by "{\it circular magnetization heat
current}, ${\bf j}^Q_M=c({\bf M}\times{\bf E})$," (${\bf M}$ is
the magnetization) that was used in [4] is wrong. In fact, the
term $({\bf M}\times{\bf E})$ is the {\it magnetization part of
the Poynting vector} [2]. Without such correction the results of
[4] contradict the third law of thermodynamics. Besides this, we
also stressed [2] that the Gaussian model is fully applicable to
ordinary superconductors, for which [4] predicts the fluctuation
correction to $\beta$ to be at least $\epsilon_F / T \sim 10^5$
times bigger than $\beta$ for noninteracting electrons. Moreover,
far above $T_c$ this huge effect was predicted to decrease as
$T^{-2}$ and, therefore, it would dominate in the wide temperature
range up to $ \sim 100 T_c$, i.e. up to the room temperatures [5].
Certainly, such huge effects are not known for ordinary
superconductors (Nb, Al, Sn, and etc), which just slightly above
$T_c$ show the same values and temperature dependencies as
nonsuperconducting metals. It is also well understood that large
thermomagnetic effects are observed in materials with small Fermi
energy, i.e. with the large particle-hole asymmetry [6,7].

In their Reply [1], Serbyn, Skvortsov, and Varlamov did not
address our general objections [2] to their Letter [4] and instead
criticized our microscopic calculations in the previous paper [3].

The key result of our work [3] is that the heat current vertex for
fluctuating Cooper pairs (the Aslamzov-Larkin block in the diagram
technique, see Fig. 1) is modified by the magnetic field,
\begin{eqnarray}
{\bf B}^h = {\omega \over 2e} B \biggl({\bf q} + {2e\over c} {\bf
A}^H \biggr),
\end{eqnarray}
where $(\omega, q)$ are the energy and momentum of Cooper pairs,
near the transition $B$ is some constant, and ${\bf A}^H$ is the
vector potential of the magnetic field.

Our opponent agree that we {\it "correctly obtain that the heat
current is proportional to the gauge-invariant momentum"} [1] (see
Eq. 2 in [1]).

At the same time, our opponents stated that we {\it "fail to
include ${\bf A}^H$ in the electric vertex and draw the diagrams,
extracting ${\bf A}^H$ from the propagators and from the heat
vertex only"} [1]. Obviously, they assumed the electric current
vertex in the magnetic field has a form
\begin{eqnarray}
{\bf B}^e = B \biggl({\bf q} + {2e\over c} {\bf A}^H \biggr).
\end{eqnarray}
They claimed that we did not take into account the second term in
this equation.

Here we clarify our calculations. Below we will show that in the
gauge we used in [3] this term does not contribute to the
thermomagnetic coefficient. In a general case, this terms gives a
gauge invariant expression for the correlator of electric and heat
currents.

Let us present electric and magnetic fields as
\begin{eqnarray}
{\bf E}= i {\Omega \over c} {\bf A}^E  \ \ \ \ \ \ {\bf H}= i[{\bf
k} \times {\bf A}^H],
\end{eqnarray}
then the thermal current in the thermomagnetic effect is
proportional to
\begin{eqnarray}
{\bf E}\times {\bf H} = \Omega {\bf A}^H ({\bf k}\cdot {\bf A}^E)
- \Omega {\bf k}( {\bf A}^E \cdot {\bf A}^H).
\end{eqnarray}
As usually, let us put ${\bf E}$ along $x$-axis and  ${\bf H}$
along $z$-axis, then ${\bf E}\times {\bf H}$ will be in the
negative direction of $y$-axis. Then Eq. 4 can be presented as
\begin{eqnarray}
[{\bf E}\times {\bf H}]_y = -\Omega {\bf A}^H_y ({\bf k}_x \cdot
{\bf A}^E_x) + \Omega {\bf k}_y( {\bf A}^E_x \cdot {\bf A}^H_x).
\end{eqnarray}

To find the thermomagnetic coefficient using the Kubo method, one
should calculate the correlator of the heat current directed along
$y$ and the electric current directed along $x$,
\begin{eqnarray}
{\bf B}^h_{y} &=& {\omega \over 2e} B \biggl({\bf q} + {2e\over c}
{\bf A}^H \biggr)_{y} =  {\omega \over 2e}B q_y + {\omega \over c}
B A^H_y \\ {\bf B}^e_x &=& B \biggl({\bf q} + {2e\over c} {\bf
A}^H \biggr)_x = B q_x + {2e\over c} B A^H_x.
\end{eqnarray}
\begin{figure}
\caption{Fluctuation AL diagrams describing the heat
current-electric current correlator in crossed electric and
magnetic fields. Wavy lines stand for the fluctuation propagators
and straight lines stand for the electron Green functions, which
form the AL blocks.}
\includegraphics{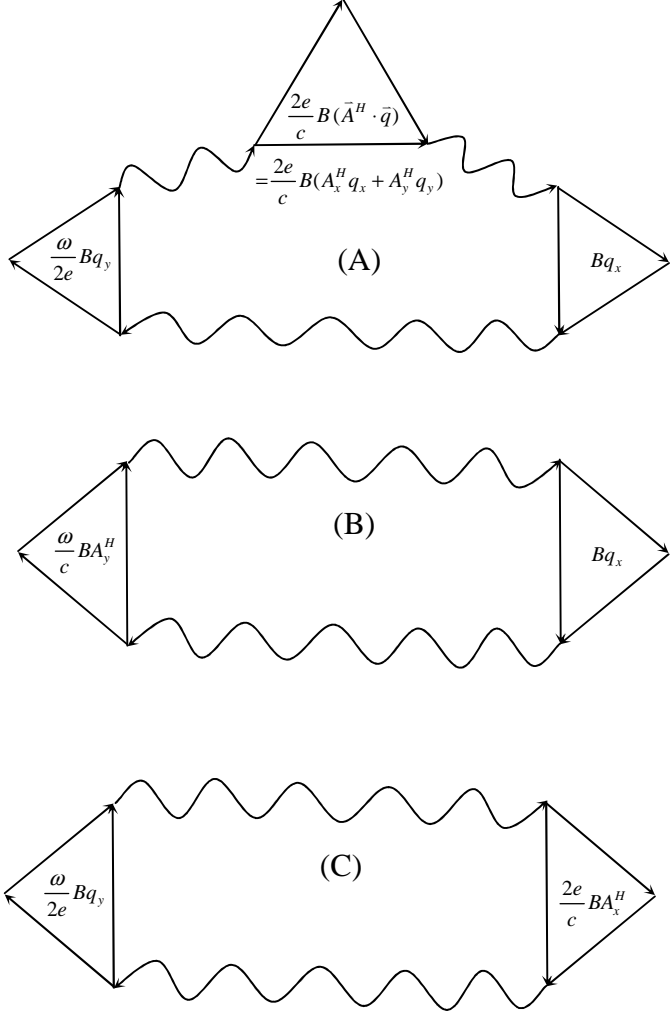}
\end{figure}

We agree with [1], that calculating thermomagnetic coefficient of
fluctuating pairs one should extract $A^H$ from the propagators of
Cooper pairs, heat current vertex, and electric current vertex.
The corresponding diagrams are presented in Fig. 1. In the diagram
(A), $A^H$ is extracted from the propagators, in the diagram (B)
it is extracted from the heat current vertex (Eg. 6), and in the
diagram (C) it is extracted from the electric current vertex (Eq.
7).

Our opponents accused us in overlooking of the diagram (C) [1]:

{\it "The relevant diagrams for thermoelectric response contain
three sources of the vector potential/magnetic-field dependence:
(A) Green functions/propagators, (B) heat current vertex, and (C)
electric current vertex. The resulting expression is gauge
invariant only if all these three sources are taken into account
in a consistent fashion and within a specific gauge. In Ref. [3]
it is claimed that the contribution (B) cancels the contribution
(A) calculated by Ussishkin in Ref. [8]. However the consistency
between calculations and gauge choices in the two parts of the
same physical quantity is not discussed. Most importantly,
contribution (C) is not even mentioned by Sergeev et al., which
makes their conclusion erroneous."}

We have a simple answer to this criticism. In [3] we used the
gauge
\begin{eqnarray}
{\bf A}^H = (0, -A^H, 0), \ \ \ \ {\bf k} = (-k, 0,0).
\end{eqnarray}
Obviously, in this gauge the diagram (C) gives zero contribution,
because  ${\bf A}^H_x =0$. This gauge is widely used for
calculations of the Nernst and Hall coefficients, as in this case
the second term in Eq. 5 is zero and $[{\bf E}\times {\bf H}]_y$
is equal to $-\Omega {\bf A}^H_y ({\bf k}_x \cdot {\bf A}^E_x)$
(see Eq. 5).

We can choose another gauge with
\begin{eqnarray}
{\bf A}^H = (A^H, 0, 0), \ \ \ \ {\bf k} = (0, -k,0).
\end{eqnarray}
In this case the diagram (B) gives zero contribution, because
${\bf A}^H_y = 0$. However, it is easy to see that now the diagram
(C) gives exactly the same contributions as the diagram (B) in the
previous gauge. Now the term $[{\bf E}\times {\bf H}]_y$ is given
by $\Omega {\bf k}_y( {\bf A}^E_x \cdot {\bf A}^H_x)$ (see Eq. 5).

Obviously, in a general case the diagrams (B) and (C) provide
thermoelectric effect, which is proportional to the gauge
invariant expression for $[{\bf E}\times {\bf H}]$ given by Eq. 5.

In conclusion, we explain with all details that in [3] we
correctly ignore the diagram (C), because in the gauge we used in
[3] the diagram (C) is zero. In any other gauge this diagram gives
nonzero contribution. The sum of the contributions of the diagrams
(B) and (C) gives the gauge-invariant term, which cancels the
contribution of the diagram (A) in zero order in the particle-hole
asymmetry (PHA). The nonzero thermomagnetic coefficient arises
only in the second order in PHA [3].

Thus, we confirm that in the Fermi liquid with the particle-hole
excitations, the thermomagnetic coefficients are always
proportional to the square of the particle-hole asymmetry.
Therefore, huge thermomagnetic effects observed in high-$T_c$
cuprates can be associated with the larger particle-hole asymmetry
due to the Fermi surface reconstruction or due to a non-Fermi
liquid state, such as the vortex liquid.

We are grateful M.N. Serbyn, M.A. Skvortsov, and A.A. Varlamov for
detailed discussion of our work [3].

\end{document}